\documentclass{aa}
\usepackage{graphics}
\usepackage{amsmath}
\usepackage{amssymb}
\newcommand{\df}{{\rm d}}
\newcommand{\e}{{\rm e}}
\newcommand{\half}{\tfrac{1}{2}}
\newcommand{\gccm}{\mathrm{\,g\,cm}^{-3}}

\begin{document}

\thesaurus{01(08.14.1; 08.05.3; 08.18.1; 02.18.8; 13.25.5)}

\title{Two Dimensional Cooling Simulations of Rotating Neutron
	Stars} 
\author{Christoph Schaab\thanks{E-mail address:
	schaab@gsm.sue.physik.uni-muenchen.de} \and
	Manfred K. Weigel}
\institute{Institut f{\"u}r theoretische Physik,
  Ludwig-Maximilians Universit{\"a}t M{\"u}nchen, Theresienstr. 37,
  D-80333 M{\"u}nchen, Germany}

\date{Received ....... / Accepted .......}

\titlerunning{Cooling of rotating neutron stars}

\maketitle

\begin{abstract}
The effect of rotation on the cooling of neutron stars is
investigated. The thermal evolution equations are solved in two
dimensions with full account of general relativistic effects. It is
found that rotation is particularly important in the early epoch when
the neutron star's interior is not yet isothermal. The polar surface
temperature is up to 63\,\% higher than the equatorial
temperature. This temperature difference might be observable if the
thermal radiation of a young, rapidly rotating neutron star is
detected. In the intermediate epoch ($10^2\lesssim t\lesssim
10^5$~yr), when the interior becomes isothermal, the polar
temperature is still up to 31\,\% higher than the equatorial
temperature. Afterwards photon surface radiation dominates the cooling,
and the surface becomes isothermal on a timescale of $\sim
10^7$~yr. Furthermore, the transition between the early and the
intermediate epochs is delayed by several hundred years. An additional
effect of rotation is the reduction of the neutrino luminosity due to
the reduction of the central density with respect to the non-rotating
case.

\end{abstract}

\section{Introduction}

In the past three decades the cooling history of neutron stars was
investigated by several authors
(e.g. \cite{Tsuruta66,Richardson82,VanRiper91b,Schaab95a,Page97a}). Recent
numerical simulations account for non-isothermal interior, as well as
for general relativistic effects. Nevertheless, as far as we know, all
investigations assumed spherical symmetry of geometry and temperature
distribution.

As it was pointed out by Miralles et~al. (1993), the effect of rapid
rotation on the cooling of neutron stars can be as important as
general relativistic effects, whereas the effect of slow rotation
should be negligible. Although the assumption of slow rotation holds
for most of the known pulsars, there exist a couple of millisecond
pulsars (\cite{Lyne96b}), for which rotation should yield a rather
different cooling behaviour. These millisecond pulsars are generally
located in binary systems. It may however be expected that young,
isolated millisecond pulsars can be detected in the near future,
too. A candidate might be the supernova remnant of SN1987A. Although
the observed neutrino burst lasting for ten seconds indicates that a
neutron star was formed in the supernova, there is no evidence for the
continued existence of it (s. \cite{Chevalier97a} for a recent
review). However, the neutron star can still be hidden by the
surrounding matter, and the continued observations might reveal a
rapidly rotating neutron star.

The aim of this letter is to study the effect of non-spherical
geometry on the cooling of neutron stars. As far as we know, this is
the first investigation of rotational effects beyond the isothermal
core ansatz (\cite{Miralles93}) and also the first completely two
dimensional simulation of neutron star cooling. The letter is
organized as follows: We first derive the general relativistic
equations of thermal evolution and describe the numerical method in
Sect. \ref{sec:eq}. In Sect. \ref{sec:res}, we apply the two
dimensional cooling code to static and rotating neutron star models
based on a relativistic equation of state including hyperonic degrees
of freedom. Finally, we summarize our conclusions and discuss further
improvements and applications of the current work in
Sect. \ref{sec:concl}.

\section{Equations and Numerical Method} \label{sec:eq}

Already a few seconds after the formation of a neutron star in a
supernova, its interior settles down into catalysed, degenerate
matter. The subsequent cooling involves only thermal processes and
does not change the space time geometry. However, the structure of a
neutron star depends on the rotational velocity, which generally
decreases as the star loose angular momentum, e.g. due to emission of
magnetic dipole radiation. Since the time scale for reaching
hydrostatic equilibrium is much smaller than the time scale for the
variation of angular velocity (s. \cite{Zeldovich71}, p. 239), one can
treat the evolution of a neutron star in quasi stationary
approximation. Though the partial transformation of rotational energy
into thermal energy may considerably change the cooling behaviour of a
neutron star (see, e.g., \cite{VanRiper91b}), and also the variation
of space time geometry might have an effect on it, we study here, as a
first step, the simplest case of constant angular velocity.

The stationary, axisymmetric, and asymptotic flat metric in quasi
isotropic coordinates reads
\begin{equation} \label{eq:cool.metric}
  \df s^2 = -\e^{2\nu}\df t^2+\e^{2\phi}(\df\varphi-N^\varphi \df t)^2 
  +\e^{2\omega}(\df r^2+r^2\df\theta^2),
\end{equation}
where the metric coefficients $g_{\mu\nu}=g_{\mu\nu}(r,\theta)$ are
functions of $r$ and $\theta$ only. The metric coefficients are
determined by the Einstein equation ($c=G=1$) ${\bf G} = 8\pi{\bf T}$,
and the energy-momentum conservation ${\bf\nabla}\cdot{\bf T}=0$. The
obtained elliptic differential equations (\cite{Bonazzola93a}) are
solved via a finite difference scheme (\cite{Schaab97a}) once, before
the cooling simulation starts.

In the case of uniform rotation, $\Omega={\rm const.}$, considered
here, the equations for thermal evolution are
(\cite{Miralles93,Schaab97a})
\begin{align}
  \partial_r \tilde h_1 + \frac{1}{r}\partial_\theta \tilde h_2 
	&= -r\e^{\phi+2\omega}\left(\frac{1}{\Gamma}\e^{2\nu}\epsilon
	+\Gamma C_{\rm V}\partial_t\tilde T\right) \label{eq:evol}\\ 
  \partial_r \tilde T &= 
	-\frac{1}{r\kappa}\e^{-\nu-\phi}\tilde h_1 \label{eq:H1}\\
  \frac{1}{r}\partial_\theta \tilde T &=
	-\frac{1}{r\kappa}\e^{-\nu-\phi}\tilde h_2 \label{eq:H2},
\end{align}
where
\begin{align}
  \Gamma &=
  \left(1-\e^{2(\phi-\nu)}\left(\Omega-N^\varphi\right)^2\right)^{-\half} \\
  \tilde h_i &= \frac{1}{\Gamma}r\e^{2\nu+\phi+\omega}h_i \\
  \tilde T   &= \frac{1}{\Gamma}\e^\nu T.
\end{align}
$\bf h$ denote the heat flux 3-vector in the comoving frame, $C_{\rm
V}$ the heat capacity, $\epsilon$ the neutrino emissivity, and
$\kappa$ the heat conductivity. The partial radial and angular
differentials are abbreviated by $\partial_r$ and $\partial_\theta$,
respectively.  Thermal equilibrium is described by $\tilde T={\rm
const.}$.

At the surface of the neutron star the heat flux $h_1$ and $h_2$ is
determined by the normal heat flux $h_{\rm N}$
\begin{align}
  h_1(r=R) = h_{\rm N}^1 &= 
    \left(1+\frac{1}{R^2}\left(\frac{\df R}{\df\theta}\right)^2
	\right)^{-\frac{1}{2}}h_{\rm N} \label{eq:gleich.cool.cond3}\\
  h_2(r=R) = h_{\rm N}^2 &= 
    -\left(1+\frac{1}{R^2}\left(\frac{\df R}{\df\theta}\right)^2
	\right)^{-\frac{1}{2}}
     \frac{1}{R}\frac{\df R}{\df\theta} h_{\rm N}, \label{eq:gleich.cool.cond4}
\end{align}
where $R(\theta)$ is the $r$-coordinate of the surface. $h_{\rm N}$ is
taken from a non-magnetic photosphere model which describes the
temperature gradient in the region between $e=10^{10}\gccm$ and the
star's surface (e.g. \cite{Gudmundson83}). In these models $h_{\rm
N}(\theta)$ depends on the temperature at the density $e=10^{10}\gccm$
and on the surface gravity
\begin{multline}
  g_{\rm s} = \Gamma\e^{-\nu-\omega}
    \left(\left(\partial_r\frac{1}{\Gamma}\e^\nu+\Gamma\e^{-\nu+2\phi}
	(\Omega-N^\varphi)\partial_r\Omega\right)^2 \right. \\
    \left.\left. +\frac{1}{r^2}\left(\partial_\theta\frac{1}{\Gamma}\e^\nu
	+\Gamma\e^{-\nu+2\phi}(\Omega-N^\varphi)\partial_\theta\Omega\right)^2
    \right)^{1/2}\right|_{\rm surface}.
\end{multline}

The parabolic differential equations obtained after inserting
Eqs. \eqref{eq:H1} and \eqref{eq:H2} into Eq. \eqref{eq:evol} are
solved via an implicit finite difference scheme by using a alternating
direction implicit method. This yields a non-linear equation system
which can be solved iteratively. The obtained linear equation systems
have tridiagonal coefficient matrices which can be inverted rather
fast.  The correctness of the two dimensional code was checked by
comparing the outcome of it with simple, analytically solvable models
and with the results of the one dimensional code described by
Schaab et~al. (1996).

\section{Results} \label{sec:res}

We consider a superfluid neutron star model basing on the relativistic
Hartree-Fock equation of state labelled RHF8 in Huber et~al. (1997), which
accounts for hyperonic degrees of freedom. The global properties of
uniformly rotating models with fixed gravitational mass
$M=1.5\,M_\odot$ and angular velocity $\Omega=0$, $0.5$, and
$0.99\,\Omega_{\rm K}$ are summarized in
Tab. \ref{tab:models}. $\Omega_{\rm K}$ denotes the maximum possible
Kepler angular velocity, above which mass shedding sets in. All models
allow for both the direct nucleon Urca and for the direct hyperon Urca
processes (cf. \cite{Prakash92}). All direct Urca processes are
suppressed by nucleon and lambda pairing below the respective critical
temperature (cf. \cite{Schaab98b}). The ingredients to the cooling
simulations are similar to those discussed by
Schaab et~al. (1996) in detail and are published on the Web
(http://www.physik.uni-muenchen.de\hspace{0pt}/sektion\hspace{0pt}%
/suessmann\hspace{0pt}/astro\hspace{0pt}/cool\hspace{0pt}/schaab.0198%
\hspace{0pt}/input.html).
\begin{table}[tbp]
\caption[]{Models of uniformly rotating neutron stars with fixed
gravitational mass $M=1.5M_\odot$ and angular velocity $\Omega=0$,
$0.5$, and $0.99\,\Omega_{\rm K}$ \label{tab:models}}
\begin{tabular}{lccc}
\hline
Model	& $\Omega=0$	& $\Omega=0.5\,\Omega_{\rm K}$ 
	& $\Omega=0.99\,\Omega_{\rm K}$	\\
\hline
$M$ [$M_\odot$]			& 1.5	& 1.5	& 1.5	\\
$M_{\rm B}$ [$M_\odot$]		& 1.694	& 1.693	& 1.672	\\
$R_{\rm eq}^\infty$ [km]	& 12.81	& 13.25	& 17.03	\\
$\Omega$ [s$^{-1}$]		& 0	& 2631	& 5208	\\
$e_{\rm c}$ [$10^{14}\gccm$]	& 11.67	& 10.84	& 7.208	\\
$n_{\rm c}$ [fm$^{-3}$]		& 0.6285& 0.5888& 0.4065\\
$g^{\rm eq}_{\rm s}$ [$10^{14}{\rm cm}^2{\rm s}^{-1}$]	
				& 1.648	& 1.459	& 0.546	\\
$g^{\rm p}_{\rm s}$ [$10^{14}{\rm cm}^2{\rm s}^{-1}$]	
				& 1.648	& 1.630	& 1.396	\\
\hline
\end{tabular}

\smallskip
Entries are the gravitational mass $M$, the baryonic mass $M_{\rm B}$,
the circumferential radius as measured by a distant observer $R_{\rm
eq}^\infty$, the angular velocity $\Omega$, the central mass density
$e_{\rm c}$, the central baryon density $n_{\rm c}$, and the surface
gravity $g_{\rm s}$.
\end{table}

\begin{figure}[tbp]
\centering\resizebox{0.7\hsize}{!}{\includegraphics{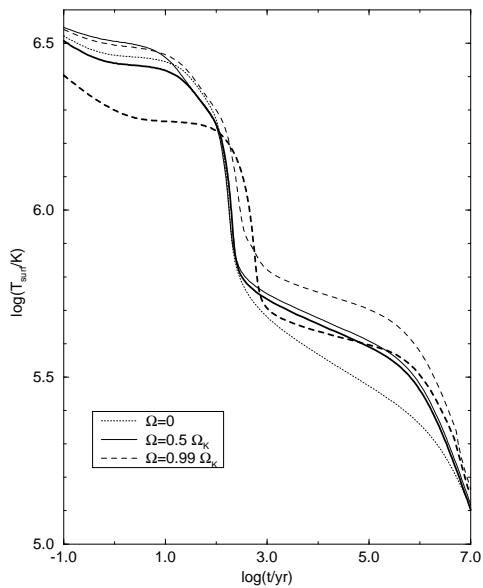}}
\caption[]{Thermal evolution of rotating, superfluid neutron
  stars. The thick (thin) curves correspond to the equatorial (polar)
  surface temperature as measured by a distant observer. The
  considered models are given in Tab. \ref{tab:models}. \label{fig:cool}}
\end{figure}
\begin{figure}[tbp]
\centering\resizebox{0.7\hsize}{!}{\rotatebox{-90}{\includegraphics{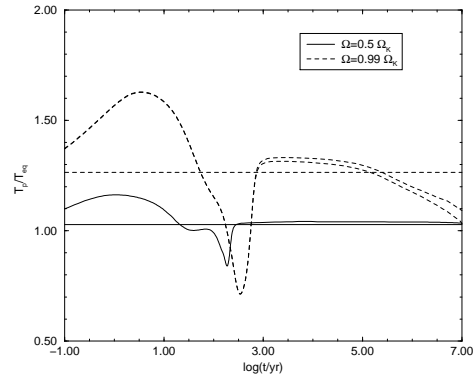}}}
\caption[]{Fraction $T_{\rm s}^{\rm p}/T_{\rm s}^{\rm eq}$ of polar
  and equatorial surface temperature as measured by a distant observer
  for the two rotating models. The corresponding relations $(g_{\rm
  s}^{\rm eq}/g_{\rm s}^{\rm p})^{1/4}$ of surface gravity are shown
  by the horizontal lines.  The thin dashed curve is calculated under
  the assumption that $T_{\rm s}\propto g_{\rm s}^{1/4}$ holds even
  for small surface temperatures. \label{fig:deviation}}
\end{figure}

Figure \ref{fig:cool} shows the evolution of the surface temperature
as measured by a distant observer. It is possible to distinguish three
epochs of evolution. In the first epoch $t\lesssim 100$~yr, large
temperature gradients occur in the interior of the neutron
star. Radial temperature gradients were already found in one
dimensional simulations (see, for example, \cite{Richardson82}). In
rotating neutron stars, azimuthal temperature gradients, which cause
transverse heat flow, $h_2\neq 0$, exist, too. The polar temperature
is by 16\,\% or 63\,\% higher than the equatorial temperature for the
models rotating with $\Omega=0.5\,\Omega_{\rm K}$ and
$0.99\,\Omega_{\rm K}$, respectively (see also
Fig. \ref{fig:deviation}). After about 100~yr the cooling wave reaches
the surface and the interior becomes isothermal, $\tilde T={\rm
const.}$ The temperature deviation between pole and equator is now
mainly caused by the smaller surface gravity $g_{\rm s}$ at the
equator, since $T_{\rm s} \propto g_{\rm s}^{1/4}$ for fixed internal
temperature. Whereas the temperature deviation is still high in the
case of nearly Kepler rotation ($\sim 31$\,\%), it almost disappears
for the model with $\Omega=0.5\,\Omega_{\rm K}$ ($\sim 4$\,\%). After
about $10^5$~yr, the photon surface radiation dominates the
cooling. The temperature distribution tends to a new equilibrium with
$\df T_{\rm m}(\theta)/\df t={\rm const.}$, on a timescale of $\sim
10^7$~yr. $T_{\rm m}(\theta)$ denotes the temperature at the inner
boundary of the photosphere at $e=10^{10}\gccm$. Since the heat
capacity is only weakly temperature dependent for $T\lesssim 10^9$~K
in the outer crust, the surface temperature tends to an isothermal
state, whereas the temperature in the outer crust varies with the
surface gravity. This behaviour is accompanied by the breakdown of the
scaling $T_{\rm s}\propto g_{\rm s}^{1/4}$ of the surface temperature
with surface gravity for small surface temperatures. However, as one
can see by comparing the thin dashed curve, which assumes $T_{\rm
s}\propto g_{\rm s}^{1/4}$ over the whole temperature range, with the
thick curve in Fig. \ref{fig:deviation}, this is only a small effect
and cannot explain the tendency of the surface temperature fraction to
unity.

Besides these effects on the angular dependency of the temperature,
rotation has also a net effect on cooling in the intermediate
epoch. This effect is caused by the reduction of the neutrino
luminosity, which sensitively depends on the central density. A
further effect of rotation on cooling of neutron stars is the
lengthening of the thermal diffusion time by several hundred
years. Additionally the drop of the surface temperature is smoother in
the case of rotation (see also \cite{Miralles93}), since the cooling
wave reaches the polar region earlier than the equator. During the
cooling wave is reaching the different surface regions, the fraction
$T_{\rm s}^{\rm p}/T_{\rm s}^{\rm eq}$ falls below unity
(s. Fig. \ref{fig:deviation}).

\section{Conclusions and Discussion} \label{sec:concl}

In this letter, we have studied the effect of non-spherical geometry
on the cooling of neutron stars. Our general finding is that rapid rotation
plays a significant role in the thermal evolution. This role is
particularly important in the early epoch, when the star's interior is
not yet isothermal. It is precisely this epoch, when even isolated
pulsars might rotate very rapidly. The angular dependent radial and
transverse heat flow cause azimuthal temperature gradients in the
interior of the star and thus also at the surface. The polar surface
temperature is by up to 63\,\% higher than the equatorial
temperature. The simulation of this non-isothermal epoch cannot be
performed with isothermal core approaches or one dimensional codes.

In the intermediate epoch, heat conduction is unimportant, since the
interior is nearly isothermal. Nevertheless two significant, though
smaller, effects remain. The first effect is the reduction of the
central density with respect to the static, non-rotating case, if one
fixes the gravitational mass or the rest mass. Since the neutrino
emissivity depends sensitively on the density, especially in the
vicinity of threshold densities for fast neutrino emission processes,
the neutrino luminosity of the star is reduced as well. The second
effect is the angular dependency of the surface temperature caused by
the angular dependent surface gravity. The polar surface temperature
is still by up to 31\,\% higher than the equatorial temperature. Both
effects could also be studied with one dimensional codes or even
within the isothermal core ansatz (see \cite{Miralles93}). Whereas
Miralles et~al. (1993) assumed that the interior is still isothermal in the
late photon cooling epoch, we find that the surface temperature
distribution itself tends to an isothermal equilibrium, which implies
that the temperature distribution in the outer crust becomes angular
dependent.

The transition between the early and the intermediate epochs turns out
to be delayed by rotation. Because of the enlargement of the crust in
the equatorial plane, the cooling wave formed in the inner region of
the star needs several hundred years longer to reach the
surface. Furthermore, the transition becomes smoother, since the
radial diffusion time it not isotropic any more (see
also \cite{Miralles93}).

The surface temperature, especially in the intermediate epoch, is
rather sensitive to micro physical parameters (superfluid energy gap,
neutrino emissivity, equation of state, etc.; see \cite{Schaab95a} for
a recent review). The comparison of the net effect with observation
might therefore be difficult. Nevertheless, the comparison of the
observed soft X-ray spectra with the theoretical obtained spectra may
reveal the angular dependency of the surface temperature due to
rotational effects. The deduction of the spectrum from the surface
temperature distribution is, unfortunately, not at all trivial in
non-spherical geometry (\cite{Cunningham75a}). So far, only the
thermal spectra of slowly rotating neutron stars ($\Omega=70.37 {\rm
s}^{-1}$ for the Vela pulsar 0833-45) are known. It may however be
expected that the thermal spectra of a young, isolated millisecond
pulsar can be detected in the near future. For example, the period
$P=1.56$~ms of the fastest pulsar B1937+21 (\cite{Backer82a}) known
today corresponds to a ratio $\Omega/ \Omega_{\rm K}=0.77$, which lies
between the respective ratios of our two rotating models.

Since we neglected some important effects, like differential rotation
and the loss of angular momentum, our work can only be understood as a
first step towards two dimensional simulation of cooling of rotating
neutron stars. Of course, the two dimensional code presented here may
also serve to investigate other non-spherically symmetric effects,
such as the effect of strong magnetic fields on heat conductivity and
neutrino emissivity. The inclusion of differential rotation and strong
magnetic fields, as well as the mentioned deduction of the photon
spectra will be addressed to future work.

\begin{acknowledgements}

We would like to thank the referee K.~A. Van Riper for his valuable
suggestions. One of us (Ch.~S.) gratefully acknowledges the Bavarian
State for financial support. Tables with detailed references to the
used ingredients and the obtained cooling tracks can be found on the
Web:
http://www.physik.uni-muenchen.de\hspace{0pt}/sektion\hspace{0pt}%
/suessmann\hspace{0pt}/astro\hspace{0pt}/cool\hspace{0pt}/schaab.0198.

\end{acknowledgements}


\end{document}